\documentclass{article}

\usepackage{arxiv}

\usepackage[utf8]{inputenc} 
\usepackage[T1]{fontenc}    
\usepackage{hyperref}       
\usepackage{url}            
\usepackage{booktabs}       
\usepackage{amsfonts}       
\usepackage{nicefrac}       
\usepackage{microtype}      
\usepackage{lipsum}
\usepackage{graphicx}
\graphicspath{ {./images/} }

\title{Toward Human-Aligned Luminance Measurement for Large-Format LED Displays}

\author{
 Xi Mou \\
  Stanford University\\
  \texttt{mmou@stanford.edu} \\
   \And
 Xiaopeng Peng \\
  Rochester Institute of Technology\\
  \texttt{xxp4248@rit.edu} \\
  \And
 Tongsheng Mou \\
  Zhejiang Smart and Health Lighting Research Center\\
  \texttt{mou@sensingm.com} \\
}

\begin{document}
\maketitle
\begin{abstract}
Direct-view LED displays are widely adopted in large-format applications due to their high luminance and reliability. However, visual comfort and accurate performance evaluation remain challenging due to the complex interaction between pixel luminance, human visual perception, and measurement artifacts. In this work, we introduce a novel 2D imaging luminance meter that replicates key optical parameters of the human eye, including entrance pupil size and angular resolution, to assess perceived pixel luminance. We report comprehensive measurements across various visual field angles and distances and establish a refined luminance metric that aligns with foveal vision standards (1/120 degree). Furthermore, a new method to quantify and mitigate stray light effects significantly improves measurement precision—reducing luminance overestimation from 7\% to 2\%. Our findings provide a foundation for optimizing LED display design for perceptual comfort and advancing standardization in pixel luminance evaluation
\end{abstract}

\section{Introduction}
Light Emitting Diode (LED) lighting technology, with its exceptional display quality and energy efficiency,  achieved a market size of 87.10 billion USD in 2023 and is expected to increase to 298.38 billion USD by 2030 \cite{fortunebiz2024, yang2025rf}. The demands for live streaming of entertainment shows, sports, advertisements, campaigns, etc. lead to the development of the large-format LED displays. Most of the past works on imaging and display system focuses on improving either the optical resolution or the spatial resolution \cite{de2016limits, peng2017randomized}. Enhancement of image qualities through machine learning algorithms \cite{peng2015image, peng2016shape} have also been studied. None of these approaches physically characterizes the image quality in terms of human eye perception.  


In this study, we introduce a novel 2D imaging luminance meter (LMD) that replicates key features of the human visual system—including photopic spectral response, entrance pupil size, and foveal angular resolution (1/120 degree)—to quantify perceived pixel luminance. Unlike traditional CCD-based luminance meters, our system enables accurate measurement of local luminance variations aligned with the observer's visual field. This approach allows for a more precise evaluation of display comfort and performance. We investigate the dependency of perceived pixel luminance on viewing distance and angular resolution, and we propose a peak luminance ratio metric based on foveal averaging. We also examine the impact of stray light on measurement accuracy and demonstrate how an optimized optical setup significantly mitigates this effect. These results provide practical guidelines for display designers and establish new benchmarks for standardizing luminance evaluation in large-format LED systems.

\begin{figure}[!t]
\centering
\includegraphics[width=3.1in]{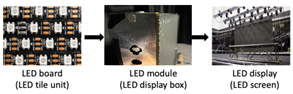}
\caption{Three stages of manufacturing the direct-viewing LED display: (a) An LED board is made from a number of discrete RGB chips; (b) An LED display box is integrated from LED boards and driver circuits; and (c) A fully assembled LED display.  }
\label{fig: fig1}
\end{figure}

\section{Related Work}

The performance evaluation of large-format LED displays has typically focused on metrics such as spatial resolution, contrast ratio, and average screen luminance. While image enhancement techniques—ranging from optical improvements \cite{peng2016randomized, peng2017randomized} to machine learning-based methods \cite{peng2021cnn, peng2022computational, peng2024learning, peng2025learning}, have advanced image clarity and uniformity, they do not address the human perception of peak brightness at the pixel level. Luminance uniformity corrections such as dot correction and tile-level compensation are widely adopted to address manufacturing variability. These methods often rely on conventional CCD-based 2D imaging luminance meters, which 

We addressed image quality issues in large format displays as well as the measurement methods of the perceived peak pixel luminance and luminance ratio of a large LED display in previous studies \cite{mou202216, wang2022measurement}. Here, we extend our discussion to the design of an optical imaging system that can simulates the optical characteristics of human eyes. We review in this section the direct-view LED technologies, the large-format LED displays, as well as human visual system.  We explain the role of measuring the perceived pixel luminance in designing the large-format LED that produces the best visual comfort. 

\textbf{Direct-view LED Displays}. The early television industry, liquid crystal displays (LCDs) dominate the market, where the liquid crystal is backlit by LED. The direct viewing LEDs do not require backlights, edge lights, and a shutter. In these displays, the LEDs directly light the display without any filters. The direct viewing LEDs panels are lighter, more reliable and energy efficient than traditional LCD panels. 
Different from the residential television displays or the small-format displays in augmented or virtual realities \cite{mou2024p, mou202438}, large-format displays are typically viewed from far away, where high luminance and high contrast are prioritized over the pixel density. Without the light cutoff by the liquid crystal layer, the luminance range of direct viewing LED displays can be as high as 4,000-10,000 cd/$\mathrm{m}^2$ \cite{schwiegerling2004field}. Displays with high luminance is especially beneficial for bright ambient illumination (e.g., daylight sports stadium). In addition, the resolution of newer generation LED is improving, with the pixel pitch  reduces from millimeters to sub-millimeters \cite{svilainis2009estimation}. However, the spacing between pixels in LED displays remain relatively large compared to LCD and OLED displays. The visual comfort is affected by the peak-to-average pixel luminance ratio, where a high ratio may lead to glare and thus eye discomfort. The viewing distance for large-format LED displays should also be far enough (1,000 times the pixel pitch of the display) that the subtended visual angle (0.057 degrees per pixel, or 17 pixels per degree) aligns with the angular resolution of the human eyes.  

\begin{figure}[!t]
\centering
\includegraphics[width=3.1in]{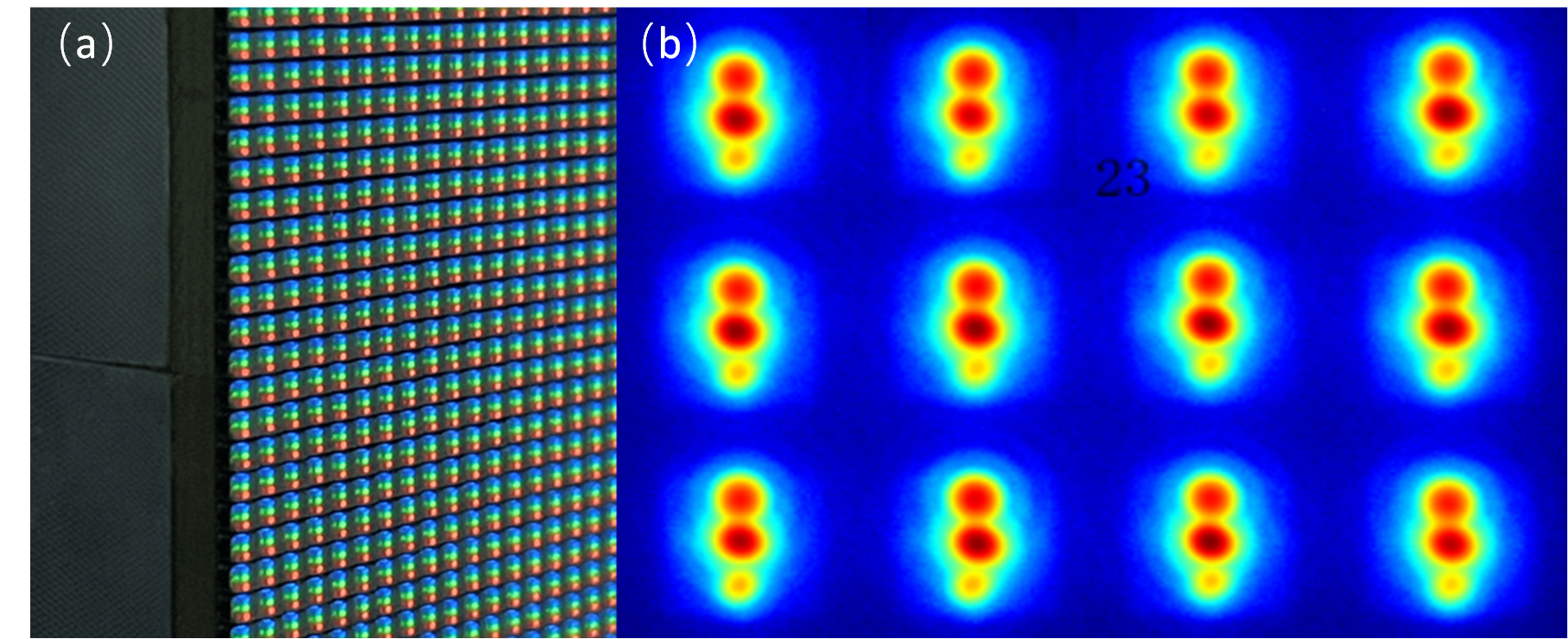}
\caption{(a) An example of LED modules showing individual pixel units. (b) An example of luminance distribution of LED pixel units in pseudo color (measurement data obtained from Hangzhou SanTest testing laboratory). }
\vspace{-1em}
\label{fig: fig2}
\end{figure}

\textbf{Large-format LED Displays}. As illustrated in Figure 1 the manufacturing stages of a direct viewing LED display, an LED screen is made of a large number of discrete RGB chips. A large number of RGB chips integrate into a substrate, creating an LED board (LED tile unit, see Fig. 1(a)). Several LED-boards are integrated with driver circuits to form an LED display box (see Fig. 1(b)). A number of display boxes then built up into the display box (see Fig. 1(c)). The input signal goes through the signal distributor to each display box to form the full-screen display. Every single chip corresponds to one pixel of the image. The pixel pitch describes the distance between the adjacent LEDs on the screen and prescribes the minimum viewing distance for an intended image.
Large-format LED display has a large pixel pitch, typically in the range of 1 mm to 3 mm. LED chips are smaller in size relative to the pixel pitch. Each pixel unit is composed of three sub-pixels of RGB-LED chips (see Fig. 2(a)). Without surface homogenization, their luminance profiles are typically ununiform.  An example of the luminance distribution of a very small section of the LED screen is shown in Fig. 2(b). The individual LED pixel unit produces high image quality such as better contrast. However the optical performance can vary due to their dependencies on the driving current and junction temperature, the non-uniform luminance and chromaticity values, as well as uneven spatial light distributions among discrete LEDs that occur in the manufacturing process and assembly \cite{ahn2013implementation}. To obtain high-quality display performance, it is necessary to correct the luminance and chromaticity values of each pixel. In the work of Svilanis  \cite{svilainis2009led}, it is demonstrated that luminance non-uniformity tends to occur at low intensities. In addition, the illuminance uniformity can be affected by the direction of individual LEDs, intensity bins, driving current, or even current distribution on PCB. The tuning for the uniformity of individual LEDs can be accomplished using dot correction \cite{schubert2006light}. A 2D imaging luminance meter is typically used to measure the pixel luminance of a LED screen, however, the angular resolution of these CCD-type imaging light measuring devices (LMDs) are not high enough to measure the pixel luminance distribution. 

\begin{figure}[!t]
\centering
\includegraphics[width=3.1in]{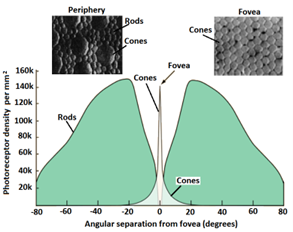}
\caption{Illustration of photoreceptors on human retina. The density curves for the rods and cones on the retina demonstrate the high density of cones in the center fovea. These cones are responsible for both color vision and the highest visual acuity. Visual inspection of fine detail involves focusing light from the details onto the center fovea region. The rods are absent from the fovea area. Their density rises to a high value and spreads over a large area of the retina at a few degrees away from the fovea region. These rods contribute to human night vision, motion detection, and peripheral visions. }
\vspace{-1em}
\label{fig: fig3}
\end{figure}

\textbf{Human Visual System}. To benchmark  the interaction between human visual system and LED displays in terms of perceived luminance, we review here the psycho-physical research on visual comfort. The density or size of photoreceptors in the retina determines the visual angular resolution of the eye. The structure and distribution of the photoreceptors on the retina is shown in Fig. 3. The small sharp area covered by the fovea determines the angle of view of our central vision. The rest of the retinal area determines the much larger angle of view of our peripheral vision. The fovea consists of a large number of cone cells, which dominates our photopic vision. This fovea region corresponds to a field of view angle of about 2 to 5 degrees. The density of cone cells decreases rapidly with the outward field of view. The cone cells vary in diameter from 1 to 4 m in the central field and increase to about 10 m in the peripheral field of view \cite{ahn2013implementation}. In addition, the image plane on our retina is semi-spherical, not flat as the CCD photoelectric sensor of a conventional camera. The eye is a variable focal length optical system, and the image distance remains constant at different viewing distances, while the visual angular resolution corresponding to the cone cells also remains constant. Beyond these classic visual psychophysics models, recent research suggests that perceptual experience of visual content can be modulated not only by retinal and cortical mechanisms but also by measurable neural intent. In parallel, studies also show that neural responses can be systematically decoded to capture affective states under visual stimulation \cite{xiao2025multi}, further underscoring how cognitive state, emotional context, and selective attention jointly influence perceived visual quality. Similarly, novel interaction paradigms \cite{yang2025rf} enables contact-free eyeglass-based control, suggest future display systems may incorporate unobtrusive, perception-aware feedback loops. Together, these advances point toward a broader design framework for large-format LED displays, where both physiological optics and neural signals inform perceived luminance and visual comfort.

Table 1 presents a side-by-side comparison of the relevant physiological and optical parameters of the human eye and those implemented in our proposed luminance meter. The system’s detector pitch, focal length, and entrance pupil diameter are selected to closely simulate the eye’s ability to resolve luminance structure within the foveal field. This design ensures that measured pixel luminance corresponds to how the human observer perceives fine-scale brightness.

To ensure that an image does not appear to be pixelated to human eyes, the viewing distance must be addressed in the design of large-format LED displays. In addition, the properties of human eyes need to be considered when designing a 2D imaging luminance meter for the perceived pixel luminance measurements. Theories that relates the human vision factor with the design of LED displays include the Watson's equation \cite{watson2016pyramid} and Kulikowski's law \cite{kulikowski1971some}. Watson’s work provides an generalization of the Ferry-Porter Law \cite{tyler1990analysis} by relating the fusion frequency with the logarithm of retinal illuminance. It provides a theoretical foundation for the understanding the interaction between LED displays and human vision at various spatial frequencies and luminance levels. The Kulikowski's law states that there is a trade-off between spatial and temporal resolution in the human visual system. Essentially, the more detailed an object is (high spatial frequency), the harder it is to detect its motion; conversely, if an object is less detailed (low spatial frequency), its motion is easier to detect.

\begin{figure}[!t]
\centering
\includegraphics[width=3.1in]{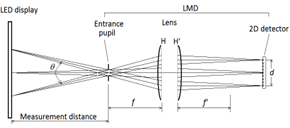}
\caption{The schematics of the proposed luminance meter, which characterizes the human eye viewing condition.}
\vspace{-1em}
\label{fig: fig4}
\end{figure}

In practical terms, this means that fine patterns or high-resolution images are not perceived as moving as smoothly as broader, less detailed patterns. This is thought to be due to the differences in how our visual system processes detail versus movement, with different types of cells in the retina and visual cortex being responsible for static detail (spatial frequency) and motion (temporal frequency). Kulikowski's Law is relevant in various fields, such as vision science, neurology, and areas of design and technology where understanding motion perception is important, like animation, film, and user interface design. Recent studies \cite{watson2016pyramid, kulikowski1971some, tyler1990analysis} indicates that by combining the Kulikowski and Watson’s work, better accuracies can be achieved.

\section{Novel Imaging Luminance Meter}
Here we introduce a novel 2D imaging luminance meter. Factors that characterize the perception of human eyes include: 
\begin{itemize}
\item 	The LMD's spectral response matches the spectral luminous efficiency function of photopic vision. To provide the necessary spectrum response V(l), a colored filter is used in conjunction with the CCD array detector and optical lens. In the measurement, the CIE-$f_1$' spectral mismatch factor is less than 0.05 \cite{iso2014characterization}.

\item The LMD remains constant in the measurement of angular resolution at any measurement distance. The LMD applies a spherical coordinate system to mimic eye rotation and eye movement. To do so, the origin of the viewing is located at the center of the eye pupil, i.e., the center of the entrance pupil of the LMD. 

\item The LMD's entrance pupil replicates the size and position of the eye's pupil have successfully demonstrated an accurate measurement for human perception \cite{wang202119, wang2022choice}. The use of varying entrance pupil sizes can have an impact on the measurement outcomes for LED spatial beams with non-uniform profiles \cite{oshima201679, janisse1973pupil}. When a display is bright, the human eye's pupil dilation occurs in the 2 to 5 mm range \cite{cherng2020background, mou2010measurement}. Similar to the eye, the LMD receives light from the screen at a specific location based on the pupil position. The distance between the pupil and the screen is used to determine the measurement distance.

\item The LMD has a sufficiently high angular resolution for averaging the perceived pixel luminance. The averaging calculation from the captured 2D image is performed to mimic the angular resolution of the human eye. The schematics of our luminance meter is shown in Figure 4, an aperture stop is set at the front focal point of the optical lens, which is equivalent to the entrance pupil [29].  Each array detector element of the LMD corresponds to a constant angular field of view for the measurement in a certain direction in front. The measurement light beam falls perpendicularly on the photosensitive surfaces of the detector elements at a small cone angle respectively, either in the center field of view or in the edge field of view, ensuring consistent response of the detector and reducing cross stray light effects. The pixel spacing on the detector corresponds to the measured angular interval. In the experiment, the focal length of the optical lens is f, and the size of the detector element is 

\end{itemize}

\begin{figure}[!t]
\centering
\includegraphics[width=3.1in]{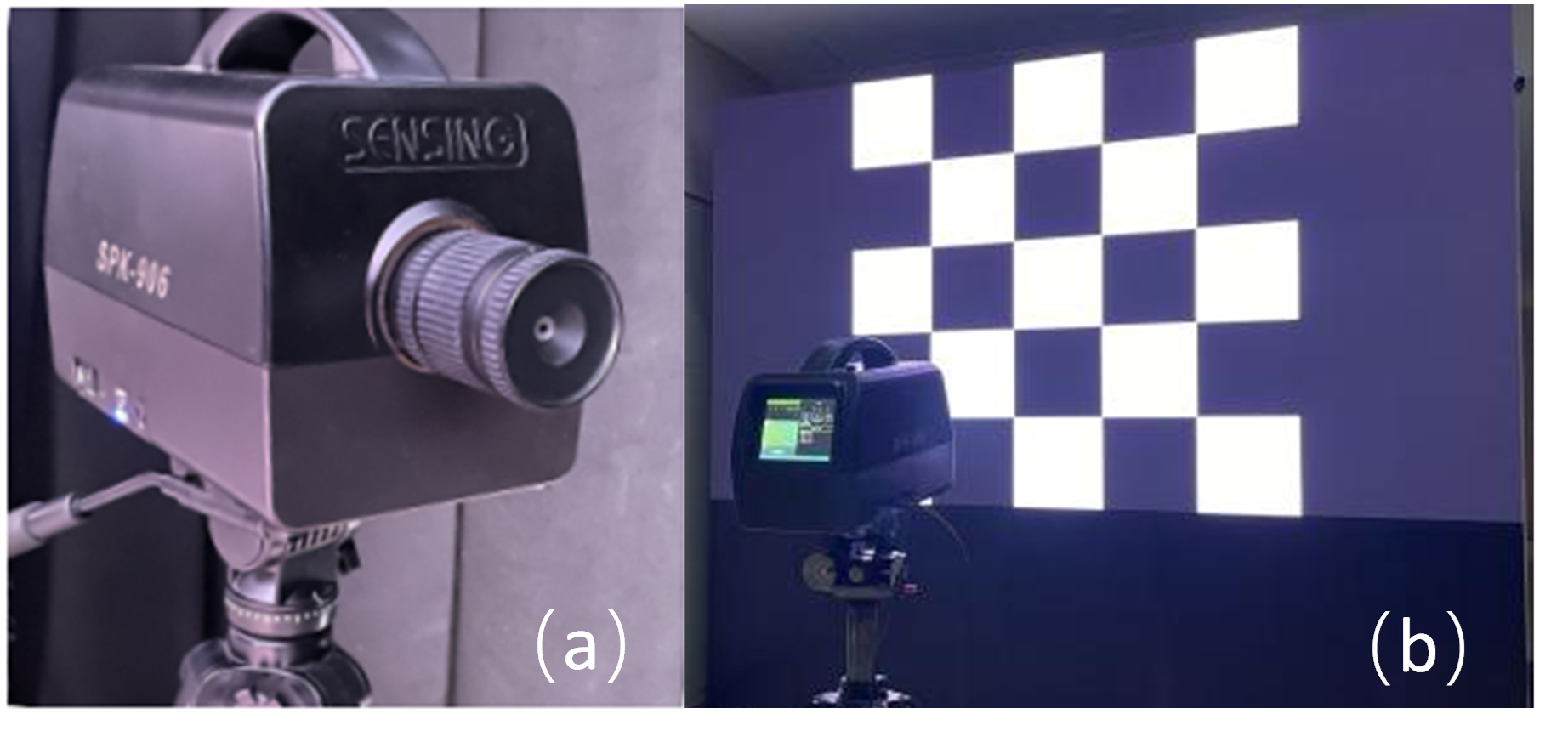}
\caption{Measurement setup (a) An example of a 2D imaging luminance meter, (b) experiment setup with a LED screen, and a 2D imaging luminance meter. Three experiments were carried out in this study. In the first experiment, the goal is to identify the LED screen luminance at different field of views (FOVs): The luminance distribution of each LED pixel on the screen on a 2-degree field of view was measured at different distances with a luminance meter detector (LMD) with a pixel angle resolution of 0.0011 degrees (909 PPD). The average luminance within different viewing angles was drawn as a function of viewing angle. And the pixel point luminance of the LED screen can be based on 1/60 degree (60 PPD).}
\vspace{-1em}
\label{fig: fig5}
\end{figure}

\begin{figure*}[t]
\centering
\includegraphics[width=7in]{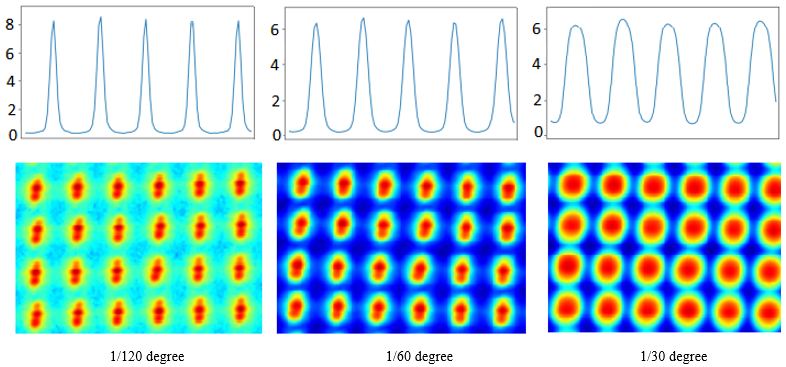}
\caption{Filtered luminance distribution with three local FOVs. (Row 1: An example of a 2D imaging luminance meter with a 4mm aperture stop located at the front focal point of the lens; Row 2: Experiment installation with a 150-inch LED screen and a 2D imaging luminance meter captured pseudo color luminance map.)}
\vspace{-1em}
\label{fig: fig6}
\end{figure*}

The schematics of our luminance meter is shown in Figure 4, an aperture stop is set at the front focal point of the optical lens, which is equivalent to the entrance pupil [29].  Each array detector element of the LMD corresponds to a constant angular field of view for the measurement in a certain direction in front. The measurement light beam falls perpendicularly on the photosensitive surfaces of the detector elements at a small cone angle respectively, either in the center field of view or in the edge field of view, ensuring consistent response of the detector and reducing cross stray light effects. The pixel spacing on the detector corresponds to the measured angular interval. In the experiment, the focal length of the optical lens is f, and the size of the detector element is $\Delta d$, so the detector element corresponds to an angle of measurement field of $(\Delta d/f)\times(180/\pi)$ degrees, which corresponds to $(\Delta d/f\times D)$ area on the LED chip at a measurement distance, $D$. The total measurement FOV can be determined by $\theta=d/f$, and the measurement area on the LED screen is covered by $(\Delta d/f\times D)$. In the measurement of perceived pixel luminance profiles, the 2D imaging luminance meter should have a high enough angular resolution that each LED pixel unit has enough detector elements to correspond. 

The measurement setup is shown in Fig. 4, where the entrance pupil position of the LMD represents the viewing point of the human eye. The measurement coordinate system of the 2D imaging LMD is determined by the spherical coordinate system of the human eyes.  For example, the 2D imaging luminance meter here has a detector element of 2.4 µm and a lens focal length of 125 mm, which corresponds to a measured angular interval of 0.0011 degrees. For a measurement distance of 1.5 meters, the measured area width on the corresponding LED pixel unit is 28.8 µm, which is approximately 52 detector photocells to correspond to the LED pixel unit with a 1.5mm pitch. Moreover, when measuring the luminance of a single pixel with a 2D imaging luminance meter, the surrounding light-emitting LED pixels can produce stray light that affects the measurement uncertainty. The effect of the surrounding stray light should be deducted from the measurement.

\section{Experiment}
The current study leverages an evolved experiment setup to elucidate the perceived luminance under varying conditions. The 2D imaging luminance meter remains central in the experiment, now enhanced with a more precise calibration to capture subtle variations in light distribution. The measurement protocol has been refined to assess the impact of ambient light conditions and viewing distance with greater accuracy. We introduce an approach for quantifying stray light effects, improving upon our original findings. This meticulous attention to experimental conditions ensures that our data more accurately reflects the complexities of visual perception in real-world LED display use.

The experiment set up is shown in Fig. 5, where we use the 2D imaging luminance meter SPK-906 (Zhejiang Sensing Optronics) as our LMD. It provide a state-of-the-art focal length of 125 mm. The meter is equipped with an entrance pupil measuring 4 mm in diameter, strategically positioned at the front focal point to optimize the accuracy of luminance readings. The detector array is composed of elements each 2.4 µm by 2.4 µm in size, facilitating high-resolution imaging. Additionally, the detector features $V(\lambda)$ correction to align its spectral responsivity with the photopic luminous efficiency function, thereby mirroring the human eye's response to light. The luminance meter is designed with programmable autofocus, allowing precise forward and backward translation, ensuring the image sharpness.
To ensure an environment conducive to accurate luminance assessment, measurements were conducted under darkroom conditions, with the floor and walls finished in black to maintain a reflectivity of less than 10
In this study, three experiments were carried out respectively for the measurement of: 1) the perceived pixel luminance; 2) perceived peak luminance ratio; and 3) stray light effect.  In the first experiment, we identify the perceived pixel luminance of the LED screen at different angular resolutions. The perceived pixel luminance is obtained by the averaging the LED pixel luminance profile over a given local FOV.  As shown in Fig. 6, the filtered luminance distribution is calculated for local FOV of 1/120 degree,1/60 degree, and 1/30 degree respectively, which affects the perceived pixel luminance. 
In regard to the eye characteristics, the photoreceptor cell of 3.5 µm in the diameter at the center vision, the eye diameter of about 24 mm and the focal length of 17 mm, and the average refractive index of 1.437 are applied.  The minimum visual field angle is calculated to be approximately 1/120 degree. For example, the element size of  2.4 µm x 2.4 µm is used in the 2D imaging LMD, which is equivalent to having more than 7 detector elements in the diameter to measure the luminance profile of the LEDs within the minimum visual field angle. The luminance distribution of each LED pixel on the screen over a 2-degree field of view was measured at different distances with the 2D luminance mete.

luminance meter detector (LMD) with a measurement angle resolution of 0.0011 degrees (909 PPD). The average luminance within different viewing angles was drawn as a function of viewing angle. And the perceived pixel luminance of the LED screen can be determined according to the perceived. visual field angles as 1/120 degree, 1/60 degree ,1/30 degree and  In the second experiment, the measurement distance (viewing distance) from the test LED screen to the LMD (light measuring device) is varied from 0.5 m to 5.0m to evaluate the change of the perceived pixel luminance with the viewing distance.

In the third experiment, the measurement distance is kept at 1.5 m. The goal of this experiment is to identify the effect of stray light from each adjacent LED on the display when measuring the perceived pixel luminance. The measurements were performed at different pixel numbers: 1, 3 by 3, 5 by 5, and 7 by 7. For example, in the 1-pixel condition, only one pixel was shown and measured by the LMD for its pixel luminance in the setting of 1/60 degree in the perceived visual field angle. in the 1-pixel condition, only one pixel was shown and measured by the LMD for its pixel luminance in the setting of 1/60 degree in the perceived visual field angle.

\begin{figure}[t]
\centering
\includegraphics[width=3.1in]{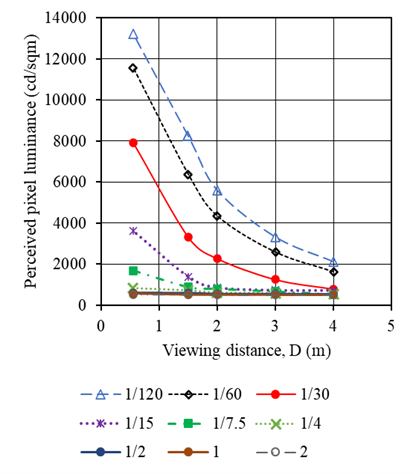}
\caption{Perceived peak pixel luminance as a function of viewing distance, D, under different viewing angles. The measurement of perceived peak pixel luminance and perceived peak pixel luminance ratio at different viewing angles as a function of viewing distance. The viewing distance is the measurement distance of the LMD. Five viewing distances include 0.55 m, 1.5 m, 2 m, 3 m, and 4 m. At each viewing distance, an angular resolution of 1/120, 1/60, 1/30, 1/15, 1/7.5, 1/4, 1/2, 1, and 2 degrees is measured. The pixel luminance of the screen viewed by the LMD changes with the viewing distances. At a particular viewing distance, the perceived pixel luminance can be varies with the viewing angular resolution.}
\vspace{-1em}
\label{fig: fig7}
\end{figure}

\begin{figure}[t]
\centering
\includegraphics[width=3.1in]{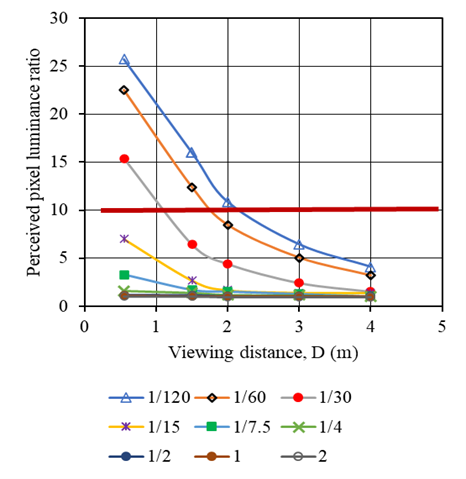}
\caption{Perceived peak pixel luminance ratio as a function of viewing distance D for different viewing angle.}
\vspace{-1em}
\label{fig: fig8}
\end{figure}

\section{Results and Discussions}
\subsection{Perceived Pixel Luminance}
The measurement of perceived pixel luminance and perceived peak pixel luminance ratio by applying different minimum perceived visual field angles as a function of viewing distance are shown in Fig. 7. The viewing distance is applied as the measurement distance of the LMD. Five viewing distances, 0.55 m, 1.5 m, 2 m, 3 m, and 4 m were included. At each viewing distance, the perceived visual field angle of 1/120, 1/60, 1/30, 1/15, 1/7.5, 1/4, 1/2, 1, and 2 in degrees, was measured. As shown in Figure 8 for the viewing distance variation, the perceived pixel luminance of the screen measured by the LMD changes significantly. This plot also indicated that at a fixed viewing distance, as the perceived visual field angle varies, the perceived pixel luminance can be very different. 

Our exploration into perceived pixel luminance of large LED displays has revealed that the interaction with human vision is more intricate than previously understood. The study breaks new ground, moving beyond initial observations to uncover a sophisticated interplay between viewing distance and perceived luminance, with angular resolution emerging as a pivotal factor. A detailed examination of stray light effects highlights the imperative for advanced calibration in luminance meters, ensuring accuracy in capturing these subtle yet critical variances.

The significance of this research lies in its illumination of the complex dynamics at play. As viewers engage with an LED screen from varying distances, the consistent average brightness belies stark variations in perceived luminance. This discrepancy becomes more pronounced as the distance decreases, intensifying the brightness contrast and potentially leading to visual discomfort. Traditional guidelines suggest a viewing distance equal to a thousand times the pixel pitch, a standard which, while generally effective, falls short in instances where perceived luminance exceeds a contrast ratio of fifteen. Such findings advocate for additional surface homogenization treatments to temper the luminance and enhance viewer comfort. 

Our methodology employed a high-resolution luminance measurement system to precisely quantify the perceived luminance of LED pixels, averaging across various visual field angles. This approach confirms that perceived luminance is contingent upon the angle of view, drawing attention to the relevance of the foveal vision standard:1/120 degree, which corresponds to a photoreceptor diameter of 3.5 µm. This angular resolution serves as a suitable benchmark for assessing pixel luminance in large displays.

\begin{figure}[t]
\centering
\includegraphics[width=3.1in]{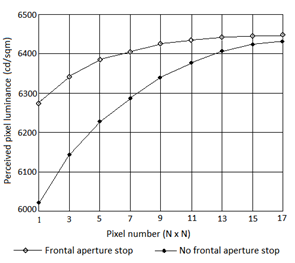}
\caption{Perceived pixel luminance in 1/60 degree as a function of varying pixel coverages ( N x N pixels, N=1 to 17) at the viewing distance of 1.5 m. The luminance at a single center pixel (with frontal aperture stop) is lower than those measured with multiple pixels are involved (no frontal aperture stop). The discrepancies may be explained by the veiling glare from the screen – a stray light effect.}
\vspace{-1em}
\label{fig: fig9}
\end{figure}

\subsection{Perceived peak pixel luminance ratio}
The perceived peak pixel luminance ratio as a function of viewing distance is shown in Fig. 8 for different viewing angles. Generally, the angular resolution of the unaided human eye is 60 PPD (1/60 degree, 1 arcmin). To keep the pixel peak luminance ratio at 10, the viewing distance should be greater than about 1.8 m. The luminance ratio is given by (Lmax-Lmin)/Lmin. We discuss the permissible limits of peak pixel luminance ratio, a metric currently undefined in the industry. Assuming a tentative threshold of ten for the visual field angle of 1/120 degree, we deduced that an effective viewing distance exceeds 2.2 meters for the studied LED display, translating to about 1500 times the pixel pitch. This ratio may vary for specific LED pixel structure and homogenized surface.

\subsection{Stray light effects}
When using 2D imaging LMD to measure the pixel luminance of the LED screen, the emission of their surrounding pixels can affect the results. In this experiment, an eye simulating 2D imaging LMD with an aperture stop placed on the front focal point is used, and the measurement light beam with a small solid angle from each viewing direction is incident perpendicularly on the detector elements, which can effectively reduce the effect of reflection on the detector.  In this study, we use black panels with different opening windows to block the light emitted from the surrounding pixels, the measurement results by this LMD and the conventional 2D camera were compared  in Fig. 9. The luminance readings of both were calibrated by a uniform light source. The measurement distance was 1.5 m from the LMD to the LED screen, and the averaged field angle for calculating the perceived luminance was 1/60 degree. The results show that for a normal imaging CCD camera, the stray light effect from adjacent pixels can reach about 7

Our investigation into the stray light effect within the LMD architecture unveils that incidental reflections from neighboring pixels, both from the imaging lens and detector surface, impart a measure of uncertainty to our luminance measurements. The newly designed imaging system mitigates this effect significantly, a testament to the advancements made in minimizing measurement errors and paving the way for future enhancements in display technologies. This holistic approach not only refines the technical aspects of LED displays but also underscores our commitment to tailoring these devices to the nuances of human vision, ultimately ensuring a comfortable and immersive viewing experience for all users.

\section{Conclusion}
In this research, a novel 2D imaging luminance meter was introduced, designed to simulate human eye perception in measuring pixel luminance distribution on large-format LED displays. Through the developed image processing pipeline, we have analyzed perceived pixel luminance and established its dependency on specific viewing conditions. The critical factors influencing visual comfort and display performance have been discerned, including the interaction between viewing distance and perceived luminance, the relevance of angular resolution to visual perception, and the impact of stray light on measurement accuracy. Our findings underscore the pivotal role of peak luminance ratio and the need for surface homogenization to mitigate the stark luminance variances perceived at closer viewing distances, which could lead to visual discomfort. It has been demonstrated that conventional guidelines for viewing distances—based on a thousand times the pixel pitch—are sometimes inadequate, particularly when the perceived luminance contrast ratio exceeds fifteen. Additionally, our study proposes a more refined benchmark for assessing pixel luminance in large displays, anchored to the foveal vision standard of 1/120 degree, and offers a methodological advance in accurately quantifying perceived luminance, thus allowing for a balanced trade-off between the optical limits of displays and human vision. The tailored luminance measurement system, which significantly reduces the stray light effect to about 2\%, represents a leap forward in precision, providing a new paradigm for evaluating LED display technology in harmony with the subtleties of human sight. This research not only paves the way for future advancements in display technologies but also emphasizes our commitment to enhancing viewer comfort, ensuring that LED screens can be enjoyed with optimal visual quality and without strain, thus enriching the user's experience.

\section{Acknowledgement}
Authors would like to extend their gratitude to the Zhejiang Smart and Health Lighting Research Center and Hangzhou SanTest Technology for providing the space and equipment to carry out this study.

\bibliography{main}  
\bibliographystyle{unsrt} 

\end{document}